\documentclass{amsart}
\usepackage{amssymb}
\usepackage{amsfonts}

\newtheorem{theorem}{Theorem}
\theoremstyle{plain}

\newtheorem{lemma}{Lemma}

\numberwithin{equation}{section}
\input{tcilatex}

\begin{document}
\title[Generalized uncertainty relations]{Generalized uncertainty relations
and \\
efficient measurements in quantum systems}
\author{V P Belavkin}
\address{Moscow Institute of Electronics and Mathematics\\
108028 Moscow USSR}
\email{vpb@maths.nott.ac.uk}
\thanks{Author acknowledges EEC support through the ATESIT project
IST-2000-29681 which allowed retyping and typesetting this paper in LaTeX. }
\date{June 20, 1975}
\subjclass{}
\keywords{Fisher Information, Uncertainty Relations, Efficient Measurements }
\dedicatory{}
\thanks{This paper is translated and typeset in LaTeX from Teoreticheskaya i
Matematichescheskaya Fizika, Vol. 26, No.3 pp. 316--329, March, 1976.}

\begin{abstract}
We consider two variants of a quantum-statistical generalization of the Cram%
\'{e}r-Rao inequality that establishes an invariant lower bound on the mean
square error of a generalized quantum measurement. The proposed complex
variant of this inequality leads to a precise formulation of a generalized
uncertainty principle for arbitrary states, in contrast to Helstrom's
variant \cite{1} in which these relations are obtained only for pure states.
A notion of canonical states is introduced and the lower mean square error
bound is found for estimating of the parameters of canonical states, in
particular, the canonical parameters of a Lie group. It is shown that these
bounds are globally attainable only for canonical states for which there
exist efficient measurements or quasimeasurements.
\end{abstract}

\maketitle

\section{Introduction}

The development in recent years of the theory of generalized quantum
measurements (see the review \cite{2} and the literature cited there) has
made it possible to introduce the concept of a quasimeasurement of
incompatible observables described by noncommuting operators and, using
this, to solve a number of problems of the quantum theory of information and
communication \cite{3,4,5,6,7,8}, give for pure states a precise formulation
of a generalised Heisenberg uncertainty principle for quantities such as,
for example, the time and energy, or phase and number of quanta \cite{9},
and to define precisely what is a measurement of the time and phase in
quantum mechanics \cite{8}, \cite{9}. In accordance with this theory, every
quantum measurement in this generalised sense is described by a positive
resolution of the identity operator $\hat{1}$ on the Hilbert space $\mathcal{%
H}$ of state-vectors $|\psi \rangle $ of the observed quantum system: 
\begin{equation}
\hat{1}=\int \Pi \left( \mathrm{d}\varkappa \right) .  \label{a}
\end{equation}%
Here $\Pi \left( \cdot \right) $ is an additive mapping (measure) on the
Borel algebra $\mathfrak{B}\left( X\right) $ of a measurable space $X\ni
\varkappa $ into the set of Hermitian-positive (i.e. nonnegative-definite
Hermitian) operators in $\mathcal{H}$. Such normalized positive measure $\Pi 
$ will be called \emph{quantum probability measure (QPM)}, or simply
quasimeasurement.\emph{\ }If $\varrho $ is a quantum state density operator,
the probability $\Pr \left( B\right) $ of an event $\varkappa \in B\,$ in
such a measurement is evaluated in accordance with the formula 
\begin{equation*}
\Pr \left( B\right) =\mathrm{Tr}\varrho \Pi \left( B\right) ,\;B\in 
\mathfrak{B}\left( X\right) 
\end{equation*}%
where $\mathrm{Tr}$ denotes the usual trace in $\mathcal{H}$. If the quantum
measure $\Pi $ in (\ref{a}) is orthogonal, $\Pi \left( A\right) \Pi \left(
C\right) =0$ for every $A\cap C=\emptyset $, then it is a projerctor-valued
measure. The generalised measurement in this case with $X=\mathbb{R}^{n}$
reduces to an ordinary measurement of the commuting self-adjoint operators 
\begin{equation}
\hat{x}^{j}=\int \varkappa ^{j}\Pi \left( \mathrm{d}\varkappa \right) ,\quad
\varkappa ^{j}\in \mathbb{R},  \label{b}
\end{equation}%
For the nonorthogonal QPM, there is no one-to-one correspondence between (%
\ref{a}) and (\ref{b}). The corresponding generalized measurements, which
are called henceforth approximate measurements of the operators (\ref{b}),
are not described as the measurements of these Hermitian operators even if
they commute, though frequently they can be described uniquely by a single
non-Hermitian (non-normal) operator (see Section \ref{EMQ}) or, more
generally, by a family of noncommuting Hermitian operators.

There is an intimate connection between the concept of a quasimeasurement as
approximate measurement and the concept of an indirect quantum measurement
(an indirect measurement is an ordinary measurement in an extended quantum
system that includes the original system as a part \cite{3}). This
connection is a simple consequence of the Naimark's well-known theorem on
the existence for every nonorthogonal QPM of an orthogonal one in an
extended Hilbert space that compresses to the original QPM on the subspace $%
\mathcal{H}$.

One of the results of this paper is to show how the concept of a generalized
measurement enables us to formulate precisely a generalised Heisenberg
uncertainty principle for quantities such as the time and energy, phase and
number of quanta, angle of rotation and angular momentum as a consequence of
a quantum Cramer-Rao type inequality for the arbitrary states. The first
members in each of these pairs -- the time, phase, and angle -- cannot, as
is well known, be described by Hermitian operators in $\mathcal{H}$, though
their measurement can be described as a statistical estimate of the
corresponding parameters of quantum states. As Helstrom has shown in \cite{9}
by means of the symmetric quantum Cram\'{e}r-Rao inequality which he
introduced in \cite{11}, the variances of the results of any measurements to
obtain such an estimate for pure states cannot be lower than a certain level
that is inversely proportional to the variances of the generators of the
unitary representations of the corresponding translation groups (i.e., the
operators of the energy, number of quanta, or angular momentum). For
example, if a pure state of a harmonic oscillator is known up to the
oscillator phase, its state-vector is unitarily equivalent to a fixed vector 
$|\psi _{0}\rangle \in \mathcal{H}$ and can be described by the family%
\begin{equation*}
\left\vert \psi _{\theta }\right\rangle =\mathrm{e}^{\mathrm{i}\theta \hat{n}%
/\hbar }|\psi _{0}\rangle ,
\end{equation*}%
where $\hat{n}$, the operator of the number of quanta, is generator of the
representation $\mathrm{e}^{\mathrm{i}\theta \hat{n}/\hbar }$ of the group
of phase translations. If a QPM $\Pi $ determines the probabilities 
\begin{equation*}
\Pr \left( \mathrm{d}\lambda |\theta \right) =\langle \psi _{\theta }|\Pi
\left( \mathrm{d}\lambda \right) \left\vert \psi _{\theta }\right\rangle ,
\end{equation*}%
on $\left[ -\pi ,\pi \right] $ such that mean value of $\lambda $ coincides
with $\theta $, 
\begin{equation*}
\mathsf{M}_{\theta }\left[ \lambda \right] :=\int \lambda \Pr \left( \mathrm{%
d}\lambda |\theta \right) =\theta ,
\end{equation*}%
it defines an unbiased estimate of the unknown value of the phase $\theta $
as the measurement result $\lambda $. The corresponding quasimeasurement,
described by the induced QPM $\Pi \left( \mathrm{d}\lambda \right) $ on $%
(-\pi ,\pi ]$, is an approximate measurement of the "phase operator" $\hat{q}%
=\int \lambda \Pi \left( \mathrm{d}\lambda \right) .$The mean quadratic
error of the measurement approximation for such $\hat{q}$ is given by the
quantum expectation $\langle \psi _{\theta }|\hat{\sigma}^{2}\left\vert \psi
_{\theta }\right\rangle $ of the positive operator%
\begin{equation*}
\hat{\sigma}_{\phi }^{2}=\int \left( \lambda -\hat{q}\right) \Pi _{\phi
}\left( \mathrm{d}\lambda \right) \left( \lambda -\hat{q}\right) ,
\end{equation*}%
and the total variance describing the estimation accuracy of $\theta $, 
\begin{equation}
R_{\theta }:=\int \left( \lambda -\theta \right) ^{2}\Pr \left( \mathrm{d}%
\varkappa |\theta \right) \equiv \mathsf{M}_{\theta }\left[ \left( \lambda
-\theta \right) ^{2}\right] ,  \label{unb}
\end{equation}%
is the sum of this and mean square distance of the operator $\hat{q}$ and $%
\theta \hat{1}$:%
\begin{equation*}
R_{\theta }=\langle \psi _{\theta }|\left( \hat{\sigma}^{2}+\left( \hat{q}%
-\theta \right) ^{2}\right) \left\vert \psi _{\theta }\right\rangle .
\end{equation*}%
The quantum Cramer-Rao inequality proves in this case that the second
variance cannot be below the level $\hbar ^{2}/4\left\langle \left( \hat{n}%
-n_{\theta }\right) ^{2}\right\rangle _{\theta }$, and thus $R_{\theta }\geq
\hbar ^{2}/4G_{\theta }$, where 
\begin{equation*}
G_{\theta }=\langle \psi _{\theta }|\left( \hat{n}-n_{\theta }\right)
^{2}\left\vert \psi _{\theta }\right\rangle =\left\langle \left( \hat{n}%
-n_{0}\right) ^{2}\right\rangle _{0},
\end{equation*}%
is Fisher information as the variance of $\hat{n}$ with $n_{\theta }=\langle
\psi _{\theta }|\hat{n}\left\vert \psi _{\theta }\right\rangle =n_{0}$. This
is Helstrom's precise formulation of the generalized Heisenberg's
uncertainty principle for the conjugate quantities $\theta $ and $\hat{n}$,
the first of which is described by a generalised measurement satisfying the
unbiased condition (\ref{unb}).

In Section \ref{Cra Rao}, we give the invariant formulation (\ref{f}) of the
Helstrom's Cram\'{e}r-Rao inequality, and we also consider another
generalization (\ref{i}) of this inequality, which in contrast to Helstrom's
can be naturally adapted to a complex situation and enables one to obtain
straightforward a multidimensional generalization of the uncertainty
relations (\ref{twoh}) for not only pure but also mixed states. We also
obtain the noncommutative generalization (\ref{twol}) of these relations for
the generators and canonical parameters of unitary representations of an
arbitrary Lie group. These generalizations are intimately related to the
canonical families of states described in Section \ref{Canon}, whose
particular role is disclosed in Section \ref{EMQ}, in which it is shown that
if the lower bounds for the mean square errors of a measurement are to be
attainable, it is necessary and sufficient that the corresponding density
operators have the canonical form (\ref{twoa}).

\section{Invariant Bounds of the Cram\'{e}r-Rao Type in quantum Statistics}

\label{Cra Rao}

\textbf{1}. Let $\left\{ \varrho _{\vartheta },\vartheta \in M\right\} $ be
a family of density operators $\varrho _{\vartheta }$ in $\mathcal{H}$ that
describe the statistical state of a quantum system as a smooth function of
unknown real parameters $\vartheta =\left( \vartheta ^{1},\dots ,\vartheta
^{m}\right) $ in a given manifold $M\subseteq \mathbb{R}^{m}$. Every
simultaneous measurement of these parameters can be described in $\mathcal{H}
$ by QPM $\Pi $ which defines a row-vector random variable $\lambda \in $ $%
\mathbb{R}^{m}$ with probability distribution $\Pr \left( \mathrm{d}\lambda
|\vartheta \right) =\mathrm{Tr}\varrho _{\vartheta }\Pi \left( \mathrm{d}%
\lambda \right) $ known up to $\vartheta $. The mean quadratic errors of the
measurement are determined by the components 
\begin{equation*}
R_{\vartheta }^{ik}=\mathsf{M}_{\vartheta }\left[ \left( \lambda
^{i}-\vartheta ^{i}\right) \left( \lambda ^{k}-\vartheta ^{k}\right) \right]
\end{equation*}%
of the covariance matrix $R_{\vartheta }=\left[ R_{\vartheta }^{ik}\right] $
by means of expressions $\mathrm{tr}C^{\intercal }R\equiv c_{ik}R^{ik}$
given by a Hermitian-positive matrix $C=\left[ c_{ik}\right] $ of the
quadratic cost form 
\begin{equation*}
c\left( \lambda ,\vartheta \right) =\left( \lambda ^{i}-\vartheta
^{i}\right) c_{ik}\left( \lambda ^{k}-\vartheta ^{k}\right)
\end{equation*}%
which plays the role of a metric tensor. Here and in what follows the
Einstein summation convention is assumed: 
\begin{equation*}
c_{ik}R_{\vartheta }^{ik}\equiv \sum_{i}\sum_{k}c_{ik}R_{\vartheta }^{ik}.
\end{equation*}%
In what follows we shall consider only those measurements that satisfy the
unbiased conditions $\mathsf{M}_{\vartheta }\left( \lambda ^{i}\right)
=\vartheta ^{i}$, under which the matrix $R_{\vartheta }$ is the covariance
matrix of the estimates $\vartheta _{i}$, and the mean square error for
fixed $R_{\vartheta }$ takes a minimal value.

Helstrom established \cite{1} for the covariance matrix $R_{\vartheta }$ of
such measurements a lower bound by using the concept of operators $\hat{g}%
_{i}$ of symmetrized logarithmic derivatives of the function $\varrho
_{\vartheta }$ with respect to $\vartheta ^{i}$. He defined these $\hat{g}%
_{i}$ by means of the equations 
\begin{equation}
\hat{g}_{i}\varrho _{\vartheta }+\varrho _{\vartheta }\hat{g}_{i}=2\partial
_{i}\varrho _{\vartheta },\;\partial _{i}:=\frac{\partial }{\partial
\vartheta ^{i}}.  \label{c}
\end{equation}%
As in the classical case \cite{11}, this bound is determined by the matrix $%
\mathsf{G}_{\vartheta }=\left[ G_{ik}\left( \vartheta \right) \right] $ of
the covariances of the logarithmic derivatives $\hat{g}_{i}=\hat{g}%
_{i}\left( \vartheta \right) $ of Eqs. (\ref{c}), defined in the symmetrized
form\ as%
\begin{equation}
G_{ik}\left( \vartheta \right) =\frac{1}{2}\left\langle \hat{g}_{i}\left(
\vartheta \right) \hat{g}_{k}\left( \vartheta \right) +\hat{g}_{k}\left(
\vartheta \right) \hat{g}_{i}\left( \vartheta \right) \right\rangle
_{\vartheta }  \label{d}
\end{equation}%
(Note that due to $\mathrm{Tr}\partial ^{i}\varrho _{\vartheta }=0$ 
\begin{equation*}
\left\langle \hat{g}_{i}\left( \vartheta \right) \right\rangle _{\vartheta
}:=\mathrm{Tr}\varrho _{\vartheta }\hat{g}_{i}\left( \vartheta \right) =0
\end{equation*}%
for all $\vartheta $). The corresponding inequality has the form 
\begin{equation}
R_{\vartheta }\geq \mathsf{G}_{\vartheta }^{-1},\quad \vartheta \in M,
\label{e}
\end{equation}%
and is understood in the sense of nonnegative definiteness of the matrix $%
\left[ R_{\vartheta }^{ik}-G_{\vartheta }^{ik}\right] $, where $G_{\vartheta
}^{ik}$ are the components of the inverse matrix $\mathsf{G}_{\vartheta
}^{-1}:G^{ij}G_{jk}=\delta _{k}^{i}$. The inequality (\ref{e}) establishing
an uncertainty relation between the variances of estimation and the
variances of the corresponding logarithmic derivatives, is a quantum analog
of the Cram\'{e}r-Rao inequality \cite{11}. The matrix $\mathsf{G}%
_{\vartheta }$ which we call symmetric quantum Fisher information, or more
fair, Helstrom information, is one of possible generalizations of classical
Fisher information. It plays the role of a metric tensor that locally
defines the geodesic distance 
\begin{equation*}
s_{\mathsf{G}}\left( \vartheta ,\vartheta +\mathrm{d}\vartheta \right)
=\left( G_{ik}\left( \vartheta \right) \mathrm{d}\vartheta ^{i}\mathrm{d}%
\vartheta ^{k}\right) ^{1/2}
\end{equation*}%
in the parameter space $M\subseteq \mathbb{R}^{m}$; this is analogous to
Fisher information distance in classical statistics.

\textbf{2.} Further, we shall consider a slightly more general situation in
which the state parameters are not the measured parameters $\vartheta ^{i}$
but local coordinates $\alpha =\left( \alpha ^{1},\ldots ,\alpha ^{n}\right) 
$ of a smooth manifold $\mathfrak{S}$ parametrizing the unknown $\varrho
=\varrho \left( \alpha \right) $. The measured parameters are assumed to be
known smooth functions $\vartheta _{i}\left( \alpha \right) $ of the unknown
parameters $\alpha ^{k}$. The corresponding generalization of Helstrom's
inequality (\ref{e}) is a lower bound for the matrix $\mathsf{R}\left(
\alpha \right) =\left[ R_{ik}\left( \alpha \right) \right] $ of the
covariances%
\begin{equation*}
R_{ik}\left( \alpha \right) =\mathsf{M}\left[ \left( \lambda _{i}-\vartheta
_{i}\right) \left( \lambda _{k}-\vartheta _{k}\right) |\alpha \right]
\end{equation*}%
of the estimates $\lambda _{i}$ in the form 
\begin{equation}
\mathsf{R}\left( \alpha \right) \geq \mathsf{D}\left( \alpha \right) \mathsf{%
G}\left( \alpha \right) ^{-1}\mathsf{D}\left( \alpha \right) ^{\intercal }.
\label{f}
\end{equation}%
that is invariant under the choice of the state coordinates $\alpha =\left(
\alpha ^{k}\right) $. Here $\mathsf{D}\left( \alpha \right) $ is the matrix $%
\left[ D_{ik}\left( \alpha \right) \right] $ of the partial derivatives $%
D_{ik}\left( \alpha \right) =\partial \vartheta _{i}/\partial \alpha ^{k}$, $%
\mathsf{D}^{\intercal }=\left[ \partial _{i}\vartheta _{k}\right] $, and $%
\mathsf{G}\left( \alpha \right) =\left[ G_{ik}\left( \alpha \right) \right] $
is the symmetric quantum Fisher information corresponding to the coordinates 
$\alpha $, that is a matrix of the covariances 
\begin{equation*}
G_{kl}\left( \alpha \right) =\mathrm{Tr}\left[ \hat{g}_{k}\left( \alpha
\right) \cdot \hat{g}_{l}\left( \alpha \right) \varrho \left( \alpha \right) %
\right] ,\;\hat{g}_{k}\cdot \hat{g}_{l}=\frac{1}{2}\left( \hat{g}_{k}\hat{g}%
_{l}+\hat{g}_{l}\hat{g}_{k}\right) ,
\end{equation*}%
of the Helstrom's logarithmic derivatives 
\begin{equation*}
\varrho \left( \alpha \right) \hat{g}_{k}+\hat{g}_{k}\varrho \left( \alpha
\right) =2\frac{\partial }{\partial \alpha ^{k}}\varrho \left( \alpha
\right) \;
\end{equation*}%
with respect to the coordinates $\alpha ^{k}$.

The inequality (\ref{f}) reduces to the classical Cram\'{e}r-Rao inequality
only when the family $\left\{ \varrho \left( \alpha \right) \right\} $ is
commutative. For noncommutative families, one can have other quantum
generalizations \cite{4}, \cite{5} of the Cram\'{e}r-Rao inequality based on
other definitions of the logarithmic derivatives; these lead to other lower
bounds for $\mathsf{R}$ that may differ from Helstrom's invariant bound $%
\mathsf{DG}^{-1}\mathsf{D}^{\intercal }$. Moreover, in the noncommutative
case it makes sense to consider also the complex-valued parameters as any
quantum state has the natural complex coordinatization $\varrho =\alpha
\alpha ^{\ast }/\mathrm{Tr}\alpha ^{\ast }\alpha $ in terms of the complex
Hilbert-Schmidt operators $\alpha $ with is the adjoint operators $\alpha
^{\ast }$as their complex conjugated. In the case of complex parameters $%
\vartheta _{i}\in \mathbb{C}$ represented by analytic functions $\vartheta
_{i}\left( \alpha ,\alpha ^{\ast }\right) $ the particular importance is
acquired by the following invariant generalization of the Cram\'{e}r-Rao
inequality based on the right and left logarithmic derivatives which were
proposed independently by the author \cite{3} and Yuen and Lax \cite{5}.

\textbf{3. }Suppose the parameters $\alpha ^{k}$ are given in pairs $\left(
\gamma ^{k},\theta ^{k}\right) \in \mathbb{R}^{2}$ which are complexified as 
$\frac{1}{2}\gamma ^{k}+\mathrm{i}\theta ^{k}\equiv \beta ^{k}$. Such
parameters $\alpha \in \mathbb{R}^{2n}$, considered as complex $n$-columns,
will often be denoted as $\beta =\left( \beta ^{k}\right) \in \mathbb{C}^{n}$%
, with $\gamma =\beta +\bar{\beta}\in \mathbb{R}^{n}$ and $\theta =\func{Im}%
\beta \in \mathbb{R}^{n}$. The partial derivatives $\partial _{k}=\partial
/\partial \beta ^{k}$, $\bar{\partial}_{k}=\partial /\partial \bar{\beta}%
^{k} $ are defined by means of the partial derivatives $\partial /\partial
\gamma ^{i}$, $\partial /\partial \theta ^{i}$ in the usual manner: 
\begin{equation*}
\frac{\partial }{\partial \beta ^{k}}=\left( \frac{\partial }{\partial
\gamma ^{k}}+\tfrac{\mathrm{i}}{2}\frac{\partial }{\partial \theta ^{k}}%
\right) ,\quad \frac{\partial }{\partial \bar{\beta}^{k}}=\left( \frac{%
\partial }{\partial \gamma ^{k}}-\tfrac{\mathrm{i}}{2}\frac{\partial }{%
\partial \theta ^{k}}\right)
\end{equation*}%
such that $\partial _{k}\beta ^{l}=\delta _{k}^{l}=\bar{\partial}_{k}\beta
^{l}$ and $\partial _{k}\bar{\beta}^{l}=0=\bar{\partial}_{k}\beta ^{l}$.

The estimated parameters $\vartheta _{i},\;i=1,\ldots ,m$ as functions of
complex $\alpha ,\bar{\alpha}$ can still be real functions of $\gamma $ and $%
\theta $. They are not assumed to be analytic with respect to $\alpha $, but
differentiable independently with respect to $\alpha $ and $\bar{\alpha}$
(e.g. given by bi-analytic functions $\vartheta _{i}\left( \alpha ,\alpha
^{\prime }\right) $ at $\alpha ^{\prime }=\bar{\alpha}$). We define the
non-Hermitian right and left logarithmic derivatives of the density operator 
$\varrho \left( \alpha ,\bar{\alpha}\right) $ by the relations 
\begin{equation}
\varrho \hat{h}_{k}=\frac{\partial \varrho }{\partial \bar{\alpha}^{k}}%
,\quad \hat{h}_{k}^{\ast }\varrho =\frac{\partial \varrho }{\partial \alpha
^{k}},\quad k=1,\ldots ,n.  \label{g}
\end{equation}%
The operators $\hat{h}_{k}=\hat{h}_{k}\left( \alpha ,\bar{\alpha}\right) $
of the right derivatives with respect to $\bar{\alpha}^{k}$ are Hermitian
conjugate at each $\alpha $ to the operators $\hat{h}_{k}^{\ast }=\hat{h}%
_{k}\left( \alpha ,\bar{\alpha}\right) ^{\ast }$ of the left derivatives
with respect to $\alpha ^{k}$, and they both have zero expectations 
\begin{equation*}
\mathrm{Tr}\hat{h}_{k}\left( \alpha ,\bar{\alpha}\right) \varrho \left(
\alpha ,\bar{\alpha}\right) =0=\mathrm{Tr}\hat{h}_{k}^{\ast }\left( \alpha ,%
\bar{\alpha}\right) \varrho \left( \alpha ,\bar{\alpha}\right) .
\end{equation*}%
The corresponding quantum Fisher information is given by the matrix $\mathsf{%
H}=\left[ H_{kl}\right] $ of covariances 
\begin{equation}
H_{kl}\left( \alpha ,\bar{\alpha}\right) =\mathrm{Tr}\,\left[ \hat{h}%
_{k}\left( \alpha ,\bar{\alpha}\right) \hat{h}_{l}\left( \alpha ,\bar{\alpha}%
\right) ^{\ast }\varrho \left( \alpha ,\bar{\alpha}\right) \right] .
\label{h}
\end{equation}%
Obviously this matrix is Hermitian-positive, and under the assumption of its
nondegeneracy it defines a positive-definite metric 
\begin{equation*}
ds_{\mathsf{H}}^{2}=H_{kl}d\bar{\alpha}^{k}d\alpha ^{l}
\end{equation*}%
in some complex domain $\mathcal{O}\subset \mathbb{C}^{n}$ of the unknowns $%
\alpha \in \mathcal{O}$.

\textbf{4. }Suppose a simultaneous measurement of the parameters $\vartheta
_{i}$ is described by a QPM $\Pi $ on $X$ that determines the estimates $%
\lambda _{i}$ of $\vartheta _{i}$ as complex-valued random variables of $%
\varkappa \in X$ with respect to the distribution 
\begin{equation*}
\mathrm{\Pr }\left[ \mathrm{d}\lambda \mid \alpha ,\;\bar{\alpha}\right] =%
\mathrm{Tr}\Pi \left( \mathrm{d}\lambda \right) \varrho \left( \alpha ,\bar{%
\alpha}\right)
\end{equation*}%
parametrized by $\alpha $.

The mean quadratic errors of the measurement are determined by the matrix $%
\mathsf{R}\left( \alpha ,\bar{\alpha}\right) =\left[ R_{ij}\left( \alpha ,%
\bar{\alpha}\right) \right] $ of covariances 
\begin{equation*}
R_{ij}\left( \alpha ,\bar{\alpha}\right) =\mathsf{M}\left[ \left( \lambda
_{i}-\vartheta _{i}\right) \left( \bar{\lambda}_{j}-\bar{\vartheta}%
_{j}\right) |\alpha ,\bar{\alpha}\right]
\end{equation*}%
which can be written as the sum $R_{ij}=\left\langle \hat{\sigma}%
_{ij}^{2}\right\rangle +Q_{ij}$\ of two kind errors. The first one is given
by the Hermitian-positive matrix of the elements%
\begin{equation*}
\left\langle \hat{\sigma}_{ij}^{2}\right\rangle \left( \alpha ,\bar{\alpha}%
\right) =\mathrm{Tr}\left[ \hat{\sigma}_{ij}^{2}\varrho \left( \alpha ,\bar{%
\alpha}\right) \right]
\end{equation*}%
as the quantum expectation of the covariance operators%
\begin{equation*}
\hat{\sigma}_{ij}^{2}=\int \left( \lambda _{i}-\hat{q}_{i}\right) \Pi \left( 
\mathrm{d}\lambda \right) \left( \lambda _{i}-\hat{q}_{j}\right)
\end{equation*}%
for the quantum estimates%
\begin{equation*}
\hat{q}_{i}=\int \lambda _{i}\Pi \left( \mathrm{d}\lambda \right)
\end{equation*}%
The second forms the mean quadratic error matrix $\mathsf{Q}=\left[ Q_{ij}%
\right] $%
\begin{equation*}
Q_{ij}\left( \alpha ,\bar{\alpha}\right) =\left\langle \left( \hat{q}%
_{i}-\vartheta _{\iota }\right) \left( \hat{q}_{j}^{\ast }-\bar{\vartheta}%
_{j}\right) \right\rangle \left( \alpha ,\bar{\alpha}\right)
\end{equation*}%
for the the operators $\hat{q}_{i}$ "estimating" the parameters $\vartheta
_{i}$.

Assuming the convergence of the integral defining $\hat{q}_{i}$, the
unbiasness condition 
\begin{equation*}
\mathsf{M}\left[ \vartheta _{i}|\alpha ,\bar{\alpha}\right] :=\int \lambda
_{i}\Pr \left( \mathrm{d}\lambda \mid \alpha ,\;\bar{\alpha}\right)
=\vartheta _{i}\left( \alpha ,\bar{\alpha}\right) ,
\end{equation*}%
for the estimates $\lambda _{i}$ can be written in the form of quantum
unbiasness%
\begin{equation*}
\left\langle \hat{q}_{i}\right\rangle \left( \alpha ,\bar{\alpha}\right) =%
\mathrm{Tr}\hat{q}_{i}\varrho \left( \alpha ,\bar{\alpha}\right) =\vartheta
_{\iota }\left( \alpha ,\bar{\alpha}\right) .
\end{equation*}%
Under this assumption the matrix $\mathsf{Q}\left( \alpha ,\bar{\alpha}%
\right) =\left[ Q_{ij}\left( \alpha ,\bar{\alpha}\right) \right] $ is the
covariance matrix of the operators $\hat{q}_{i}$, and as it is shown in the
Appendix, it has lower bound $\mathsf{Q}\geq \mathsf{DH}^{-1}\mathsf{D}%
^{\dagger }$, and therefore 
\begin{equation}
\mathsf{R}\left( \alpha ,\bar{\alpha}\right) \geq \mathsf{D}\left( \alpha ,%
\bar{\alpha}\right) \mathsf{H}\left( \alpha ,\bar{\alpha}\right) ^{-1}%
\mathsf{D}\left( \alpha ,\bar{\alpha}\right) ^{\dagger },  \label{i}
\end{equation}%
where $\mathsf{D}=\mathsf{D}\left( \alpha ,\bar{\alpha}\right) $, as in (\ref%
{f}), is the matrix $\left[ d_{ik}\right] $ of the derivatives $\partial
\vartheta _{i}/\partial \alpha ^{k}$, and $\mathsf{D}^{\dagger }=\left[ 
\overline{\partial _{k}\vartheta _{i}}\right] $ is the Hermitian adjoint
matrix.

As we shall see, even in the real case $\vartheta ^{i}=\bar{\vartheta}^{i}$,
the bound (\ref{i}) may lead to a lower bound that differs from Helstrom's
bound (\ref{f}). We shall say that (\ref{i}) is the right lower bound.
Besides this bound, we can consider other bounds, for example, the
\textquotedblleft left\textquotedblright\ bound, which is based on the left
logarithmic derivatives with respect to $\bar{\alpha}$. All these bounds are
proved in the same way as (\ref{i}) see the Appendix. Note that the right
bound in (\ref{i}) is invariant under the change of variables $\left( \alpha
^{k}\right) \mapsto \left( \vartheta _{i}\right) $ by replacing the
derivatives with respect to $\alpha ^{k}$ by derivatives with respect to the
new variables $\vartheta _{i}=\vartheta _{i}\left( \alpha ,\bar{\alpha}%
\right) $ only under the analyticity condition $\partial \vartheta
_{i}/\partial \bar{\alpha}^{k}=0$ of the transforming functions $\vartheta
_{i}\left( \alpha ,\bar{\alpha}\right) =\vartheta _{i}\left( \alpha \right) $
and the condition of nondegeneracy of the matrix of the derivatives $%
\partial \vartheta _{i}/\partial \alpha ^{k}$. Therefore, the inequality (%
\ref{i}) and its noninvariant form $\mathsf{R}\geq \mathsf{H}^{-1}$ are not
equivalent unless not only the nondegeneracy of the matrix $\mathsf{D}$ but
also the analyticity condition $\partial \vartheta _{i}/\partial \bar{\alpha}%
^{k}=0$ (i.e., the condition that the functions $\vartheta _{i}\left( \alpha
,\bar{\alpha}\right) $ are independent of $\bar{\alpha}$) hold.

\section{Canonical States and Uncertainty Relations}

\label{Canon}

In classical mathematical statistics, a particular role is played by
canonical, or exponential, families of probability distributions, for which
the Cram\'{e}r-Rao bound is attainable for a special choice of the
parameters $\vartheta $. In Section \ref{EMQ}, we shall show that in quantum
statistics an analogous role is played by the density operators of the form 
\begin{equation}
\varrho \left( \beta ,\bar{\beta}\right) =\chi \left( \beta ,\bar{\beta}%
\right) ^{-1}\mathrm{e}^{\beta ^{k}\hat{x}_{k}^{\ast }}\varrho _{0}\mathrm{e}%
^{\bar{\beta}^{k}\hat{x}_{k}},  \label{twoa}
\end{equation}%
where $\hat{x}_{k},\;k=1,\ldots ,n$ are linearly independent operators in $%
\mathcal{H}$, which may be non-Hermitian: $\hat{x}_{k}^{\ast }\neq \hat{x}%
_{k}$, and even need not commute with the conjugates $\hat{x}_{i}\hat{x}%
_{k}^{\ast }\neq \hat{x}_{k}^{\ast }\hat{x}_{i}$. We shall assume that the
generating function 
\begin{equation}
\chi \left( \beta ,\bar{\beta}\right) =\mathrm{Tr}\varrho _{0}\mathrm{e}^{%
\bar{\beta}^{k}\hat{x}_{k}}\mathrm{e}^{\beta _{k}\hat{x}_{k}^{\ast }},
\label{twob}
\end{equation}%
of the moments of these operators in the state $\varrho =\varrho _{0}$ is
defined in an open neighborhood of the origin $\beta =0$ of the complex
space $\mathbb{C}^{n}$ with finite first and second moments 
\begin{equation*}
\frac{\partial }{\partial \bar{\beta}_{i}}\chi |_{\beta =0}=0=\frac{\partial 
}{\partial \beta _{k}}\chi |_{\beta =0},\;\left\langle \hat{x}_{i}\hat{x}%
_{k}^{\ast }\right\rangle _{0}=\frac{\partial }{\partial \bar{\beta}_{i}}%
\frac{\partial _{k}}{\partial \beta _{k}}\ln \chi |_{\beta =0}
\end{equation*}%
(the operators $\hat{x}_{k}$ in (\ref{twoa}) can always be chosen to have
zero expectations $\left\langle \hat{x}_{i}\right\rangle _{0}=\partial
_{i}\ln \chi |_{\beta =0}=0$ in the state $\varrho _{0}$). We shall call
that the family of density operators (\ref{twoa}) canonical, with the
parameters $\beta ^{k}$ canonically conjugate to the quantum variables $\hat{%
x}_{k}$. In contrast to the classical case, even for selfadjoint $\hat{x}%
_{k} $ one can meaningfully consider complex values of the conjugate
parameters $\beta ^{k}$.

Particular interest attaches to the case, which does not have a classical
analog, of the canonical states (\ref{twoa}) when $\beta ^{k}$ are
imaginary, $\beta ^{k}=i\theta ^{k}$, and $\hat{x}_{k}$ are selfadjoint, $%
\hat{x}_{k}=\hat{s}_{k}=\hat{x}_{k}^{\ast }$. The parameters $\theta
_{k}=\hbar \theta ^{k}$ ($\hbar $ is Planck's constant) then take the
dimension and meaning of the classical variables which are dynamically
conjugate to their shift generators $\hat{s}_{k}$. For example, \cite{9}, if 
$\hat{s}$ is the Hamiltonian, then $\theta $ is the time, if $\hat{s}$ is
the momentum then $\theta $ is the position, and if $\hat{s}$ is the number
of quanta, or angular momentum, then $\theta $ is the phase, or polar
coordinate. For $\beta ^{k}=\mathrm{i}\hbar ^{-1}\theta _{k}\equiv \beta
_{\theta }^{k}$ the canonical states $\varrho _{\theta }=\rho \left( \beta
_{\theta },\bar{\beta}_{\theta }\right) $ (\ref{twoa}) become unitary
equivalent 
\begin{equation}
\varrho _{\theta }=\mathrm{e}^{\mathrm{i}\theta ^{k}\hat{s}_{k}/\hbar
}\varrho _{0}\mathrm{e}^{-\mathrm{i}\theta ^{k}\hat{s}_{k}/\hbar }\quad
\label{twoc}
\end{equation}%
to the state $\varrho _{0}=\varrho \left( 0,0\right) $ corresponding to the
zero value $\theta =0$.

Now we shall see that the inequality (\ref{i}) with $\theta _{k}=\hbar \func{%
Im}\beta ^{k}$ applied to the canonical family (\ref{twoc}) with commuting $%
\hat{s}_{k}=\hat{s}_{k}^{\ast }$ immediately provides the precise
formulation of a generalised Heisenberg uncertainty principle for an
unbiased estimation of $\theta $. In this the right and left logarithmic
derivatives with respect to $\bar{\beta}$ and $\beta $ for the family (\ref%
{twoa}) are equal to the symmetric logarithmic derivatives $\hat{g}%
_{k}\left( \gamma \right) =\hat{s}_{k}-\mu _{k}\left( \gamma \right) $ with
respect to $\gamma =\beta +\bar{\beta}$: 
\begin{equation*}
\hat{h}_{k}=\hat{s}_{k}-\frac{\partial }{\partial \bar{\beta}_{i}}\chi
\left( \beta +\bar{\beta}\right) =\hat{g}_{k}=\hat{s}_{k}-\frac{\partial }{%
\partial \beta _{k}}\chi \left( \beta +\bar{\beta}\right) =\hat{h}_{k}^{\ast
}.
\end{equation*}%
This implies that the Fisher informations $\mathsf{H}\left( \beta ,\bar{\beta%
}\right) $ and $\mathsf{G}\left( x\right) $ coincide with the covariance
matrix $\mathsf{S}\left( \beta +\bar{\beta}\right) $ of the commutative
family $\hat{s}_{k}$ given at the state (\ref{twoa}) by 
\begin{equation}
S_{ik}=\chi \left( \beta +\bar{\beta}\right) ^{-1}\left\langle \left( \hat{s}%
_{i}-\mu _{i}\right) \mathrm{e}^{\left( \beta +\bar{\beta}\right) ^{j}\hat{s}%
_{j}}\left( \hat{s}_{k}-\mu _{k}\right) \right\rangle _{0}.  \label{twoi}
\end{equation}%
While the complex quantum Cramer Rao bound (\ref{i}) for the unbiased
estimation of real parameters $\vartheta _{i}\left( \gamma \right) $ with $%
\gamma ^{k}=\beta ^{k}+\bar{\beta}^{k}$ coincides in this case with the
Helstrom's invariant bound (\ref{f}), it also gives immediately the
uncertainty relation 
\begin{equation}
\mathsf{R}_{\theta }\geq \frac{1}{4}\hbar ^{2}\mathsf{S}_{0}^{-1},\quad 
\mathsf{S}_{0}=\mathsf{S}\left( 0\right)  \label{twoh}
\end{equation}%
for the unbiased estimation of $\theta _{i}=\hbar \theta ^{i}$ based on the
imaginary parts $\theta ^{i}=\func{Im}\beta ^{i}$ for the canonical
coordinates $\beta $ with the fixed $\gamma =0$.

Indeed, setting $\theta _{i}=\hbar \func{Im}\beta ^{i}$ such that $\partial
\theta _{i}/\partial \beta ^{k}=\hbar \delta _{ik}/2\mathrm{i}$, we obtain
from (\ref{i}) the generalised Heisenberg uncertainty relation in the form 
\begin{equation*}
\mathsf{R}\left( \beta ,\bar{\beta}\right) \geq \frac{1}{4}\hbar ^{2}\mathsf{%
S}^{-1}\left( \beta +\bar{\beta}\right) .
\end{equation*}%
Here $\mathsf{R}$ is the mean quadratic error matrix 
\begin{equation*}
\mathsf{R}=\mathsf{M}\left[ \left( \lambda _{i}-\hbar \func{Im}\beta
^{i}\right) \left( \lambda _{j}-\hbar \func{Im}\beta ^{j}\right) |\beta ,%
\bar{\beta}\right]
\end{equation*}%
of unbiased estimates $\gamma _{i}$ and $\mathsf{S}$ is the matrix of the
covariances (\ref{twoi}) defining the uncertainty relation (\ref{twoh}) at $%
\gamma =\beta +\bar{\beta}=0$.

The uncertainty relation (\ref{twoh}) acquires the following matrix meaning:
The covariance matrix $\mathsf{R}_{\theta }=\mathsf{R}\left( \beta _{\theta
},\bar{\beta}_{\theta }\right) $ of the unbiased estimates for the canonical
parameters $\theta _{i}$ of the translation group represented in $\mathcal{H}
$ by the unitary transformations (\ref{twoc}) with the selfadjoint
generators $\hat{s}_{k}$ is in the canonical uncertainty relation with the
covariance matrix of these generators, 
\begin{equation*}
S_{ik}\left( \theta \right) :=\mathrm{Tr}\varrho _{\theta }\hat{s}_{i}\hat{s}%
_{k}=\mathrm{Tr}\varrho _{0}\hat{s}_{i}\hat{s}_{k}\equiv S_{ik}\left(
0\right) ,
\end{equation*}%
in the initial (and any other transformed) state $\varrho _{0}=\varrho
\left( 0\right) $.

This uncertainly relation holds for all commuting Hermitian operators $\hat{s%
}_{i}$, not only for those like momenta which have dynamically conjugate
observables $\hat{s}_{i}$. Helstrom derived this generalized uncertainty
relation (\ref{twoh}) in one dimensional version from his bound for the
particular case of pure states $\varrho _{0}=\left\vert \varphi
_{0}\right\rangle \left\langle \varphi _{0}\right\vert $ \cite{9}. However
for this purpose the symmetric inequality (\ref{e}) is inappropriate, and
this is why his derivation involved so complicated matrix elements
calculations.

The uncertainty relations naturally correspond to not symmetric but
antisymmetric logarithmic derivatives, defined as the Hermitian solutions $%
\hat{p}_{k}=\hat{p}_{k}\left( \theta \right) $ of the von Neumann equations 
\begin{equation*}
\left[ \varrho _{\theta },\hat{p}_{k}\right] :=\varrho _{\theta }\hat{p}_{k}-%
\hat{p}_{k}\varrho _{\theta }=\frac{\hbar }{\mathrm{i}}\frac{\partial }{%
\partial \theta _{k}}\varrho _{\theta }
\end{equation*}%
For the canonical family (\ref{twoc}) we have the solutions $\hat{p}%
_{k}\left( \theta \right) =\allowbreak \hat{s}_{k}$ which are uniquely
defined by the condition $\mathrm{Tr}\varrho _{\theta }\hat{p}_{k}\left(
\theta \right) =0$. Assuming that the solutions $\hat{p}_{k}\left( \theta
\right) $ exist for an arbitrary parametric family $\varrho _{\theta }$, one
can derive the generalized uncertainty relation for the covariance matrix $%
\mathsf{R}_{\theta }$ of the unbiased estimates 
\begin{equation*}
\mathsf{M}_{\theta }\left[ \theta _{i}\right] =\left\langle \hat{q}%
_{i}\right\rangle _{\theta }=\theta _{i},\quad \hat{q}_{i}=\int \theta
_{i}\Pi \left( \mathrm{d}\theta \right) ,\quad
\end{equation*}%
in terms of the new quantum Fisher information matrix $\mathsf{S}_{\theta }=%
\left[ S_{ik}\left( \theta \right) \right] $ given by the symmetric
covariances 
\begin{equation*}
S_{ik}\left( \theta \right) =\mathrm{Tr}\hat{p}_{i}\left( \theta \right)
\cdot \hat{p}_{k}\left( \theta \right) \varrho _{\theta }.
\end{equation*}%
It simply follows form of matrix inequality%
\begin{equation*}
\mathsf{R}_{\theta }\geq \mathsf{Q}_{\theta }\geq \frac{\hbar ^{2}}{4}%
\mathsf{S}_{\theta }^{-1},
\end{equation*}%
where $\mathsf{Q}_{\theta }=\left[ Q_{ik}\left( \theta \right) \right] $ is
the matrix of covariances 
\begin{equation*}
Q_{ik}\left( \theta \right) =\left\langle \left( \hat{q}_{i}-\theta
_{i}\right) \left( \hat{q}_{k}-\theta _{k}\right) \right\rangle _{\theta }
\end{equation*}%
with $\mathsf{R}_{\theta }-\mathsf{Q}_{\theta }=\int \left[ \left( \lambda
_{i}-\hat{q}_{i}\right) \Pi \left( \mathrm{d}\lambda \right) \left( \lambda
_{k}-\hat{q}_{k}\right) \right] \geq 0$.

Indeed, due to the unbiasness $\left\langle \hat{q}\right\rangle _{\theta
}=\theta $ we have \emph{mean canonical commutation relations} 
\begin{equation*}
\left\langle \left[ \hat{q}_{i},\hat{p}_{k}\left( \theta \right) \right]
\right\rangle _{\theta }=\mathrm{Tr}\hat{q}_{i}\left[ \hat{p}_{k}\left(
\theta \right) ,\varrho _{\theta }\right] =\mathrm{i}\hbar \frac{\partial }{%
\partial \theta _{k}}\left\langle \hat{q}_{i}\right\rangle _{\theta }=%
\mathrm{i}\hbar \delta _{ik}.
\end{equation*}%
From this and $\left\langle \left[ \hat{q},\hat{p}\right] \right\rangle
_{\theta }=2\func{Im}\left\langle \tilde{q}\hat{p}\right\rangle _{\theta }$,
where $\tilde{q}=\hat{q}-\theta $, $\tilde{p}=\hat{p}-\mu $, we derive $%
\mathsf{Q}\geq \hbar ^{2}$\textsf{$S$}$^{-1}/4$ by Schwarz inequality and $%
\left\vert \left\langle \tilde{q}\tilde{p}\right\rangle _{\theta
}\right\vert \geq \left\vert \func{Im}\left\langle \tilde{q}\tilde{p}%
\right\rangle _{\theta }\right\vert $: 
\begin{equation*}
\left\langle \tilde{q}^{2}\right\rangle _{\theta }\left\langle \tilde{p}%
^{2}\right\rangle _{\theta }\geq \left\vert \left\langle \tilde{q}\tilde{p}%
\right\rangle _{\theta }\right\vert ^{2}\geq \frac{1}{4}\left\vert
\left\langle \left[ \hat{q},\tilde{p}\right] \right\rangle _{\theta
}\right\vert ^{2}=\left( \frac{\hbar }{2}\right) ^{2}.
\end{equation*}

Note that Heisenberg's uncertainty principle is usually proved only for a
single state $\varrho =\varrho _{0}$ in the form of the Robertson inequality 
$\mathsf{R}_{0}\geq \hbar ^{2}\mathsf{S}_{0}^{-1}/4$ for the variances $%
\mathsf{R}_{0}$ and $\mathsf{S}_{0}$ of the dynamically conjugate variables
described by the canonical operators $\hat{q}_{i}$ and $\hat{p}_{k}$ in $%
\mathcal{H}$ which satisfy the exact canonical commutation relations 
\begin{equation*}
\left[ \hat{q}_{i},\hat{q}_{k}\right] =0,\quad \left[ \hat{q}_{i},\hat{p}_{k}%
\right] =\mathrm{i}\hbar \delta _{ik}I,\quad \left[ \hat{p}_{i},\hat{p}_{k}%
\right] =0\text{.}
\end{equation*}%
A more precise matrix multidimensional generalization of the Robertson
inequality in terms of the covariances of estimates of an arbitrary family
of noncommuting operators is proposed in \cite{7}. Note that Robertson
inequality implies the uncertainty relation 
\begin{equation*}
\mathsf{R}_{\theta }\geq \hbar ^{2}\mathsf{S}_{0}^{-1}/4,\quad \mathsf{S}%
_{0}=\left[ \mathrm{Tr}\hat{s}_{i}\hat{s}_{k}\varrho _{0}\right] =\mathsf{S}%
_{\theta }
\end{equation*}%
for the unbiased measurements of the unknown expectations $\theta
_{i}=\left\langle \hat{q}_{i}\right\rangle _{\theta }$ in the canonical
states (\ref{twoc}) with $\int \lambda _{i}\Pi \left( \mathrm{d}\lambda
\right) =\hat{q}_{i}$, where $\hat{q}_{i}$ satisfy the canonical commutation
relations with the canonically conjugated $\hat{s}_{k}=\hat{p}_{k}$. In this
case the unbiasness 
\begin{equation*}
\mathsf{M}_{\theta }\left[ \theta \right] =\left\langle \hat{q}\right\rangle
_{\theta }=\mathrm{Tr}\varrho _{0}\hat{q}\left( \theta \right) =\left\langle 
\hat{q}\right\rangle _{0}+\theta =\theta ,
\end{equation*}%
simply means that $\left\langle \hat{q}_{i}\right\rangle _{0}=0$ for the
state $\varrho _{0}$ as%
\begin{equation*}
\hat{q}\left( \theta \right) =\mathrm{e}^{-\mathrm{i}\theta _{k}\hat{s}^{k}}%
\hat{q}\mathrm{e}^{\mathrm{i}\theta _{k}\hat{s}^{k}}=\hat{q}+\theta
\end{equation*}%
Every such unbiased measurement has the variance $\mathsf{R}_{\theta }\geq 
\mathsf{Q}_{\theta }$, and among such measurements there is an optimal one
corresponding to $\mathsf{R}_{\theta }=\mathsf{Q}_{\theta }$. It is realized
by the direct measurement of all $\hat{q}_{i}$ described by the orthogonal
spectral measure $\Pi \left( \mathrm{d}\lambda \right) =E\left( \mathrm{d}%
\lambda \right) $ of the commutative family $\hat{q}_{i}=\int \lambda
_{i}E\left( \mathrm{d}\lambda \right) $. Note that in this case $\hat{p}%
\left( \theta \right) =\hat{p}$, and both $\mathsf{S}_{\theta }=\mathsf{S}%
_{0}$ and $\mathsf{Q}_{\theta }=\mathsf{Q}_{0}$ do not depend on $\theta $
in any state $\varrho =\varrho _{\theta }$.

Our analysis extends the Heisenberg uncertainty principle to any unbiased
measurement satisfying $\left\langle \hat{q}\right\rangle _{\theta }=\theta $%
. Note that without unbiasness the uncertanicy relation doesn't hold for
such dynamically conjugate variables as polar coordinate described by the
bounded selfadjoint operator $-\pi \hat{1}\leq \hat{q}\leq \pi \hat{1}$ and
the discrete angular momentum $\hat{s}$. In this case one can find a state $%
\varrho _{0}$ (e.g. the eigen state of angular momentum for which the
uncertancy relation is obviously not true as $\mathsf{S}_{0}=0$ and $\mathsf{%
Q}_{0}\leq \pi ^{2}$). There is no good operator $\hat{q}$ in $\mathcal{H}$
satisfying the unbiasness condition $\left\langle \hat{q}\right\rangle
_{\theta }=\theta $.

{}We now consider the general case of the non-commuting generators $\hat{x}%
_{k}$ in the canonical family (\ref{twoa}). Differentiating (\ref{twoa})
with respect to $\bar{\beta}^{k}$ and comparing the result with (\ref{g}),
we obtain 
\begin{equation}
\hat{h}_{k}=\mathrm{e}^{-\bar{\beta}^{k}\hat{x}_{k}}\chi \frac{\partial }{%
\partial \bar{\beta}^{k}}\chi ^{-1}\mathrm{e}^{\bar{\beta}^{k}\hat{x}_{k}}=%
\hat{x}_{k}\left( \bar{\beta}\right) -\mu _{k},  \label{twod}
\end{equation}%
where $\hat{x}_{k}\left( \bar{\beta}\right) =\mathrm{e}^{-\bar{\beta}^{k}%
\hat{x}_{k}}\frac{\partial }{\partial \bar{\beta}^{k}}\mathrm{e}^{\bar{\beta}%
^{k}\hat{x}_{k}}$, and $\mu _{k}=\mu _{k}\left( \beta ,\bar{\beta}\right) $
is the expectation value of $\hat{x}_{k}\left( \bar{\beta}\right) $ at the
state $\varrho =\varrho \left( \beta ,\bar{\beta}\right) $: 
\begin{equation*}
\mu _{k}=\mathrm{Tr}\hat{x}_{k}\left( \bar{\beta}\right) \varrho \left(
\beta ,\bar{\beta}\right) =\frac{\partial }{\partial \bar{\beta}^{k}}\ln
\chi \left( \beta ,\bar{\beta}\right) .
\end{equation*}%
The right Fisher information matrix (\ref{h}) is therefore the matrix of the
covariances 
\begin{equation}
h_{ik}=\mathrm{Tr}\left( \hat{x}_{i}\left( \bar{\beta}\right) -\mu
_{i}\right) \left( \hat{x}_{k}\left( \bar{\beta}\right) -\mu _{k}\right)
^{\ast }\varrho \left( \beta ,\bar{\beta}\right) =\frac{\partial ^{2}\ln
\chi }{\partial \bar{\beta}^{i}\partial \beta ^{k}}\left( \beta ,\bar{\beta}%
\right)  \label{twoe}
\end{equation}%
of the operators $\hat{x}_{k}\left( \bar{\beta}\right) $ depending
analytically on $\bar{\beta}$ (but with not necessarily analytic
expectations $x_{k}$ at $\varrho \left( \beta ,\bar{\beta}\right) $). The
inequality (\ref{i}) in the neighborhood of the point $\beta =0$ can
therefore be expressed in the form of the uncertainty relation 
\begin{equation}
\mathsf{R}\left( \beta ,\bar{\beta}\right) \gtrsim \mathsf{D}\left( \beta ,%
\bar{\beta}\right) \mathsf{S}\left( \beta ,\bar{\beta}\right) ^{-1}\mathsf{D}%
\left( \beta ,\bar{\beta}\right) ^{\dagger },  \label{twof}
\end{equation}%
which establishes an inverse proportionality between the matrix $\mathsf{S}=%
\left[ S_{ik}\left( \beta ,\bar{\beta}\right) \right] $ of the covariances 
\begin{equation}
S_{ik}=\mathrm{Tr}\varrho \left( \beta ,\bar{\beta}\right) \left( \hat{x}%
_{i}-\mu _{i}\right) \left( \hat{x}_{k}-\mu _{k}\right) ^{\ast }
\label{twog}
\end{equation}%
for the operators $\hat{x}_{k}=\hat{x}_{k}\left( 0\right) $ with the
expectations $\mu _{k}=\mathrm{Tr}\hat{x}_{k}\varrho \left( \beta ,\bar{\beta%
}\right) $ and the covariance matrix $\mathsf{R}\left( \beta ,\bar{\beta}%
\right) $ of the estimates $\lambda _{i}$ for the functions $\vartheta
_{i}\left( \beta ,\bar{\beta}\right) $ of the canonical parameters $\beta
^{k}$.

Let us consider the case when the operators $\hat{x}_{k}$ are the generators
of a Lie algebra. Suppose the operators $\hat{x}_{k}$ satisfy a Lie algebra
commutation relations 
\begin{equation}
\hat{x}_{i}\hat{x}_{k}-\hat{x}_{k}\hat{x}_{i}=C_{ik}^{j}\hat{x}_{j},
\label{twoj}
\end{equation}%
where $C_{ik}^{j}$ are the structure constants. In this case, the operators $%
\hat{x}_{i}\left( \bar{\beta}\right) $ in \ref{twod}) are linear
combinations of the generators $\hat{x}_{i}=\hat{x}_{i}\left( 0\right) $ 
\cite{12}: 
\begin{equation}
\hat{x}_{i}\left( \bar{\beta}\right) =\mathsf{K}^{-1}\left( \bar{\beta}%
\right) _{i}^{j}\hat{x}_{j},  \label{twok}
\end{equation}%
where $\mathsf{K}\left( \bar{\beta}\right) =\bar{\beta}^{k}\mathsf{C}%
_{k}\left( \mathrm{e}^{\bar{\beta}^{k}\mathsf{C}_{k}}-\mathsf{I}\right)
^{-1} $ is an $n\times n$ matrix which exists in, at least, a certain
neighborhood $\mathcal{O}\subset \mathbb{C}^{n}$ of the origin $\beta =0$,
and $\mathsf{C}_{k}=\left[ C_{ik}^{j}\right] $ are the generators of the
adjoint matrix representation 
\begin{equation*}
\mathsf{C}_{i}\mathsf{C}_{k}-\mathsf{C}_{k}\mathsf{C}_{i}=C_{ik}^{j}\mathsf{C%
}_{j}
\end{equation*}%
of the commutation relations (\ref{twoj}). Expressing the covariance matrix $%
\mathsf{H}$ of the operators (\ref{twok}) in terms of the covariances (\ref%
{twog}) of the generators $\hat{x}_{i}$, we obtain in place of \ref{twof})
the exact inequality 
\begin{equation}
\mathsf{R}\left( \beta ,\bar{\beta}\right) \geq \left( \mathsf{DK}^{\dagger }%
\mathsf{S}^{-1}\mathsf{KD}^{\dagger }\right) \left( \beta ,\bar{\beta}%
\right) .  \label{twol}
\end{equation}

In the case (\ref{twoc}), the family $\varrho _{\theta }$ is unitarily
homogeneous with respect to the Lie group having Hermitian generators $\hat{x%
}_{k}=\hat{s}_{k}=\hat{x}_{k}^{\ast }$ and canonical parameters $\theta _{i}$%
. As in the case of (\ref{twoh}), we obtain a generalized uncertainty
relation 
\begin{equation}
\mathsf{R}_{\theta }\geq \frac{\hbar ^{2}}{4}\mathsf{K}_{\theta }^{\intercal
}\mathsf{S}_{0}^{-1}\mathsf{K}_{\theta },  \label{twom}
\end{equation}%
where $\mathsf{K}_{\theta }=\mathrm{i}\theta _{k}\mathsf{C}^{k}\left( 
\mathrm{e}^{\mathrm{i}\theta _{k}\mathsf{C}^{k}}-\mathbf{1}\right) ^{-1}$
and $\mathsf{C}^{k}=\hbar ^{-1}\mathsf{C}_{k}$. In the domain $\Theta
\subseteq \mathbb{R}^{n}$ of convergence of the series 
\begin{equation*}
\left( \mathsf{I}-\mathrm{e}^{\mathrm{i}\theta _{k}\mathsf{C}^{k}}\right)
^{-1}=\sum_{m=1}^{\infty }\mathrm{e}^{\mathrm{i}m\theta _{k}\mathsf{C}%
^{k}},\;\theta \in \Theta ,
\end{equation*}%
the inequality (\ref{twom}) determines the lower bound of the mean quadratic
error of measurement of the canonical parameters for the unitary
representation $\mathrm{e}^{\mathrm{i}\theta _{k}\hat{s}^{k}}$ of the Lie
group generated by the selfadjoint $\hat{s}^{k}=\hbar ^{-1}\hat{s}_{k}$.

\section{Efficient Measurements and Quasimeasurements}

\label{EMQ}

\textbf{1.} In classical statistics, estimates whose covariance matrix
attains the minimal value, transforming the Cram\'{e}r-Rao inequality
locally or globally into an equality, are said to be efficient (locally or
globally, respectively). In quantum statistics, because of the nonunique
generalization of the Cram\'{e}r-Rao inequality, the concept of efficiency,
introduced by analogy with the classical concept, loses its universality,
and the definitions of locally efficient estimates \cite{1}, \cite{4}, \cite%
{5} based on the different variants of this generalization are not
equivalent. Therefore, we shall distinguish efficient measurements (or
estimates), for which the invariant Helstrom's bound (\ref{f}) is attained,
from efficient measurements corresponding to the right bound(\ref{i}),
calling the former Helstrom efficient and the latter right efficient. As we
shall show here, the concept of right efficiency is more universal:
Measurements that are globally Helstrom efficient are also right efficient,
but not vice versa. We show first that Helstrom efficient estimates exist
globally for the canonical families of density operators (\ref{twoa}) if the
operators $\hat{x}_{k}$ are commuting self-adjoint operators $\hat{s}_{k}$,
and the estimated parameters $\vartheta \left( \gamma \right) $ are taken to
be their expectations%
\begin{equation}
\vartheta _{i}\left( \gamma \right) =\mathrm{Tr}\hat{s}_{i}\varrho \left(
\gamma \right) =\mu _{k}\left( \gamma \right)  \label{threeb}
\end{equation}%
as the derivatives $\mu _{k}=\partial \ln \chi /\partial \gamma ^{k}$ for
the moment generating function $\chi \left( \gamma \right) =\mathrm{Tr}%
\varrho _{0}\mathrm{e}^{\gamma ^{k}\hat{s}_{k}}$ of the canonical states 
\begin{equation}
\varrho \left( \gamma \right) =\chi ^{-1}\left( \gamma \right) \mathrm{e}%
^{\gamma ^{k}\hat{s}_{k}/2}\varrho _{0}\mathrm{e}^{\gamma ^{k}\hat{s}_{k}/2}
\label{threea}
\end{equation}%
corresponding to zero imaginary parts $\func{Im}\beta ^{k}=0$ in (\ref{twoa}%
) with $\chi \left( \beta ,\bar{\beta}\right) =\chi \left( \beta +\bar{\beta}%
\right) $. Differentiating the operator-function (\ref{threea}) we find the
symmetrized logarithmic derivatives $\hat{g}_{k}=\hat{s}_{k}-\mu _{k}$ with
respect to $\gamma ^{k}$. Thus, the symmetric Fisher information (\ref{d})
in this case is the matrix of covariances 
\begin{equation}
S_{ik}=\mathrm{Tr}\varrho \left( \gamma \right) \left( \hat{s}_{i}-\mu
_{i}\right) \left( \hat{s}_{k}-\mu _{k}\right) =\frac{\partial ^{2}\ln \chi 
}{\partial \gamma ^{i}\partial \gamma ^{k}}.  \label{threec}
\end{equation}%
for the operators $\hat{s}_{k}$. However these covariances as the second
derivatives of $\ln \chi $ are the derivatives $\partial \mu _{i}/\partial
\gamma ^{k}=\partial \mu _{k}/\partial \gamma ^{i}$ of (\ref{threeb}). That
defines the matrix $\mathsf{D}=\left[ \partial \vartheta _{i}/\partial
\gamma ^{k}\right] $ in (\ref{f}) as 
\begin{equation*}
\mathsf{D}\left( \gamma \right) =\left[ \partial \mu _{i}\left( \gamma
\right) /\partial \gamma ^{k}\right] =\mathsf{S}\left( \gamma \right) .
\end{equation*}%
The inequality (\ref{f}) therefore takes the form $\mathsf{Q}\left( \gamma
\right) \geq \mathsf{S}\left( \gamma \right) $, i.e. $\left[ \mathsf{Q}_{ik}-%
\mathsf{S}_{ik}\right] \geq 0$, where $\mathsf{Q}\left( \gamma \right) =%
\mathsf{R}\left( \gamma \right) $ is the covariance matrix of the operators $%
\hat{q}=\hat{s}$ realizing the unbiased estimates by the joint measurement
of $\hat{s}_{i}$. One can take the spectral QPM $\Pi $ of the family $\hat{s}%
_{i}=\int \varkappa _{i}\Pi \left( \mathrm{d}\varkappa \right) $ and define
these estimates as spectral values $\varkappa _{k}$ for $\hat{s}_{k}$. The
covariance matrix $\mathsf{R}\left( \gamma \right) $ of such estimates
obviously achieves its minimal value 
\begin{equation*}
\mathsf{R}=\mathsf{M}_{\vartheta }\left[ \left( \lambda _{i}-\vartheta
_{i}\right) \left( \lambda _{k}-\vartheta _{k}\right) \right] =\mathsf{M}%
_{\vartheta }\left[ \left( \varkappa _{i}-\mu _{i}\right) \left( \varkappa
_{k}-\mu _{k}\right) \right] =\mathsf{S}.
\end{equation*}%
Thus, for the canonical families (\ref{threea}) with commuting self-adjoint $%
\hat{s}_{k}$ there exists a Helstrom-efficient estimation $\lambda
=\varkappa $ of the functions (\ref{threeb}) defined by the canonical
parameters $\mu _{k}$, and this is realized by an a simultaneous measurement
of the commuting observables $\hat{s}_{k}$. The domain of this efficiency
obviously coincides with the domain $\mathcal{O}\subset \mathbb{R}^{n}$ in
which $\chi \left( \gamma \right) <\infty $ is twice differentiable. It can
be shown that the opposite assertion holds in the following sense.

Suppose that the estimates $\lambda _{i}$ (i.e., the results of a
measurement) have, in a certain domain, differentiable mean values $%
\vartheta _{i}\left( \alpha \right) $ and the covariances $R_{ik}\left(
\alpha \right) $, and suppose the matrices $\mathsf{R}=\left[ R_{ik}\left(
\alpha \right) \right] $ and $\mathsf{D}=\left[ \partial \vartheta
_{i}/\partial \alpha ^{k}\right] $ satisfy the following regularity
conditions 
\begin{equation}
\frac{\partial }{\partial \alpha ^{i}}\left( \mathsf{R}^{-1}\mathsf{D}%
\right) _{k}^{j}=\frac{\partial }{\partial \alpha ^{k}}\left( \mathsf{R}^{-1}%
\mathsf{D}\right) _{i}^{j}  \label{threed}
\end{equation}%
(which are trivial in one-dimensional case). Then one can introduce the
canonical parameters $\gamma ^{k}$ by setting $\gamma ^{k}\left( \alpha
_{0}\right) =0$ for an $\alpha _{0}$ at which $\vartheta \left( \alpha
_{0}\right) =0$.

It is readily verified that for a family of density operators $\varrho
\left( \gamma \right) $ of the canonical form (\ref{threea}) the regularity
conditions are satisfied for the efficient measurement of $\vartheta
_{k}=\mu _{k}\left( \gamma \right) $ as in this case 
\begin{equation*}
\mathsf{R}\left( \gamma \right) =\mathsf{S}\left( \gamma \right) ,\quad 
\mathsf{D}\left( \gamma \right) =\mathsf{S}\left( \gamma \right)
\end{equation*}%
and therefore $\left( \mathsf{R}^{-1}\mathsf{D}\right) =\mathsf{I}$. The
proof of the opposite assertion, that if the regularity conditions are
satisfied, global Helstrom efficiency holds only for the canonical families %
\ref{threea}), is given in the Appendix for the more general complex
situation. Thus,

\begin{theorem}
\label{Tone}Under the above regularity condition the inequality (\ref{f})
becomes an equality in the domain $\mathcal{O}\subset \mathbb{R}^{n}$ iff
the density operators $\varrho \left( \alpha \right) $ have the canonical
form (\ref{threea}), where $\hat{s}_{k},\;k=1,\ldots ,n$, are Hermitian
commuting operators in $\mathcal{H}$, and the canonical coordinates $\gamma $
are functions of the parameters $\alpha $ defined by the equations 
\begin{equation*}
\frac{\partial }{\partial \gamma ^{k}}\ln \chi \left( \gamma \right)
=\vartheta _{k}\left( \alpha \right) ,\;k=1,\ldots ,n.
\end{equation*}%
The optimal estimation in this case reduces to the measurement of the
Hermitian operators $\hat{s}_{k}$ described by their joint spectral
resolution of identity, and the minimal mean square error is determined by
the matrix of their covariances (\ref{threec}).
\end{theorem}

\textbf{2.} Suppose that in a domain $\mathcal{O}\subset \mathbb{C}^{n}$ of
some complex coordinates $\alpha =\left( \alpha ^{k}\right) $ the unbiased
estimates $\lambda _{k}$ have mathematical expectations $\vartheta
_{k}\left( \alpha \right) $ and covariances $R_{ik}\left( \alpha ,\bar{\alpha%
}\right) $ satisfying the regularity conditions (\ref{threed}) 
\begin{equation}
\frac{\partial }{\partial \alpha ^{i}}\left( \mathsf{R}^{-1}\mathsf{D}%
\right) _{k}^{j}=\frac{\partial }{\partial \alpha ^{k}}\left( \mathsf{R}^{-1}%
\mathsf{D}\right) _{i}^{j},\quad \frac{\partial }{\partial \bar{\alpha}^{k}}%
\mathsf{R}^{-1}\mathsf{D}=0.  \label{threee}
\end{equation}%
(which simply means in one-dimensional case the analyticity $\partial 
\mathsf{R}^{-1}\mathsf{D}/\partial \bar{\alpha}=0$). Then, as in the real
case, one can introduce the canonically conjugate parameters $\beta
^{k}=\beta ^{k}\left( \alpha \right) $ as analytic functions satisfying the
the equations%
\begin{equation*}
\partial \beta ^{i}/\partial \alpha ^{k}=\left( \mathsf{R}^{-1}\mathsf{D}%
\right) _{k}^{i},\;\;\beta ^{k}\left( \alpha _{0}\right) =0.
\end{equation*}%
and the functions $\beta ^{k}\left( \alpha \right) $ are analytic by virtue
of the condition (\ref{threee}).

\begin{theorem}
\label{Ttwo}Under the above formulated regularity conditions, the inequality
(\ref{h}) becomes an equality if and only if the family $\left\{ \varrho
\left( \alpha ,\bar{\alpha}\right) ,\alpha \in \mathcal{O}\right\} $ has the
canonical form (\ref{twoa}), where $\varrho _{0}=\varrho \left( 0,0\right) $%
, the operators $\hat{x}_{k},\;k=1,\ldots ,n$, have simultaneously in $%
\mathcal{H}$ the right eigen QPM 
\begin{equation}
\hat{1}=\int \Pi \left( \mathrm{d}\varkappa \right) ,\quad \hat{x}_{k}\Pi
\left( \mathrm{d}\varkappa \right) =\varkappa _{k}\Pi \left( \mathrm{d}%
\varkappa \right) ,\quad \varkappa =\left( \varkappa _{1},\ldots \varkappa
_{n}\right) \in \mathbb{C}^{n}\text{,}  \label{threef}
\end{equation}%
and the canonical parameters $\beta ^{k},\;k=1,\ldots ,n$ are defined by the
equations 
\begin{equation}
\frac{\partial \ln \chi \left( \beta ,\bar{\beta}\right) }{\partial \bar{%
\beta}^{k}}=\vartheta _{k}\left( \alpha ,\bar{\alpha}\right) ,\quad \alpha
\in O.
\end{equation}%
The optimal estimation in this case reduces to a quasimeasurement of the
non-Hermitian operators $\hat{x}_{k}$ described by the resolution of the
identity (\ref{threef}), and the minimal mean square error is determined by
the matrix of the covariances 
\begin{equation}
H_{ik}=\mathrm{Tr}\varrho \left( \hat{x}_{i}-\vartheta _{i}\right) \left( 
\hat{x}_{k}-\vartheta _{k}\right) ^{\ast }.  \label{threeg}
\end{equation}
\end{theorem}

The sufficiency is proved as in Section \ref{Cra Rao}. Using the invariance
of the right bound (\ref{i}) under the analytic transformations $\alpha
\mapsto \beta $, we choose as the displaced $\alpha ^{k}$ determining this
bound the canonical parameters $\beta ^{k}$ of the family of density
operators (\ref{twoa}). The elements $\partial \vartheta _{i}/\partial \beta
^{k}$ of the matrix $\mathsf{D}$ with allowance for $\vartheta _{\iota
}=\partial \ln \chi /\partial \bar{\beta}^{i}$ then coincide with the
elements (\ref{twoe}) of the matrix $\mathsf{H}$. Since the operators $\hat{x%
}_{k}$ commute in accordance with (\ref{threef}), 
\begin{equation*}
\hat{x}_{i}\hat{x}_{k}=\int \varkappa _{i}\varkappa _{k}\Pi \left( \mathrm{d}%
\varkappa \right) =\hat{x}_{k}\hat{x}_{i},
\end{equation*}%
we have $\vartheta _{k}=\mu _{k},\,\mathsf{H}=\mathsf{S}$, where $\mu _{k}$
are the mathematical expectations of $\hat{x}_{k}$ and $\mathsf{S}$ is the
covariance matrix (\ref{twog}) of these operators. Therefore, the inequality
(\ref{i}) takes the form $\mathsf{R}\geq \mathsf{S}$. It remains to show
that the measurement described by the resolution of the identity (\ref%
{threef}) leads to an estimation for which $\mathsf{R}=\mathsf{S}$ even in
the case when the operators $\hat{x}_{k}$ do not commute with their
Hermitian conjugates: $\hat{x}_{i}^{\ast }\hat{x}_{k}\neq \hat{x}_{k}^{\ast }%
\hat{x}_{i}$ (which is the case for a nonorthogonal resolution (\ref{threef}%
)). For this, it is sufficient to take into account the representation 
\begin{equation}
\hat{x}_{i}=\int \varkappa _{i}\Pi \left( \mathrm{d}\varkappa \right) ,\quad 
\hat{x}_{i}\hat{x}_{k}^{\ast }=\int \varkappa _{i}\bar{\varkappa}_{k}\Pi
\left( \mathrm{d}\varkappa \right) ,  \label{threeh}
\end{equation}%
obtained by integrating the equations in (\ref{threef}) $x\in \mathbb{C}^{n}$
and also the conjugate equation $\Pi \left( \mathrm{d}\varkappa \right) \hat{%
x}_{k}^{\ast }=\bar{\varkappa}_{k}\Pi \left( \mathrm{d}\varkappa \right) $.
Because of (\ref{threeh}), the covariances 
\begin{equation}
R_{ik}=\int \left( \varkappa _{i}-\vartheta _{i}\right) \left( \bar{\varkappa%
}_{k}-\bar{\vartheta}_{k}\right) \mathrm{Tr}\varrho \Pi \left( \mathrm{d}%
\varkappa \right)  \label{threei}
\end{equation}%
of the estimates $\lambda _{k}=\varkappa _{k}$ obtained on the basis of the
quasimeasurement of the operators $\hat{q}_{k}=\hat{x}_{k}$ coincide with
the covariance $H_{ik}$ of these operators, which proves that this
generalized measurement is efficient for the density operators (\ref{twoa}).
The proof of the opposite assertions of Theorem 2 follows from the very
derivation of the inequality (\ref{i}) and is given in the Appendix.

\textbf{3. }Thus, the condition of (right) efficiency requires the existence
of commuting operators that have a joint right spectral resolution and play
the role of sufficient statistics, which we call right-efficient. At the
same time, it is sufficient to restrict the study of these operators to the
minimal subspace generated by the domains $\varrho \left( \beta ,\bar{\beta}%
\right) \mathcal{H}$ with density operators $\varrho \left( \beta ,\bar{\beta%
}\right) $ for $\beta \in \mathcal{O}$. further, if one considers only real
values of the parameters $\vartheta _{k}\left( \beta ,\bar{\beta}\right) $,
the optimal estimation can be described by non-Hermitian and noncommuting
(with the conjugate) operators of right-efficient statistics and is not
therefore Helstrom efficient. However, estimates that are Helstrom efficient
correspond, in accordance with Theorem \ref{Tone}, to the special case of
right efficiency for which the operators $\hat{x}_{k}$ are Hermitian. If the
operators $\hat{x}_{k}$ in (\ref{twoa}) are not Hermitian but commute with
the Hermitian conjugates, the right efficient estimates also coincide with
the complexified Helstrom efficient estimates However, the commutativity $%
\hat{x}_{k}\hat{x}_{i}^{\ast }=\hat{x}_{i}^{\ast }\hat{x}_{k}$ need not hold.

\textbf{Example.} Suppose $\hat{x}_{k}=\varphi _{k}\left( \hat{a}\right) $,
where $\varphi _{k}$ are entire functions $\mathbb{C}^{r}\rightarrow \mathbb{%
C},\;\hat{a}=\left( \hat{a}_{i},\ldots ,\hat{a}_{r}\right) $ are boson
annihilation operators satisfying the commutation relations 
\begin{equation*}
\left[ \hat{a}_{i},\hat{a}_{j}\right] =0,\quad \left[ \hat{a}_{j},\hat{a}%
_{i}^{\ast }\right] =\delta _{ij}\hat{1}.
\end{equation*}%
It is well known that the operators $\hat{a}$ have right eigenvectors $%
\left\vert \alpha \right\rangle \in \mathcal{H},\alpha \in \mathbb{C}^{r}$,
that define a nonorthogonal resolution of the identity: 
\begin{equation*}
\hat{1}=\int \left\vert \alpha \right\rangle \left\langle \alpha \right\vert
\prod_{i=1}^{r}\frac{1}{\pi }\mathrm{d}\func{Re}\alpha _{i}\mathrm{d}\func{Im%
}\alpha _{i},\quad \hat{a}_{i}\left\vert \alpha \right\rangle =\alpha
_{i}\left\vert \alpha \right\rangle .
\end{equation*}%
Obviously, the operators $\hat{x}=\varphi \left( \hat{a}\right) $ also have
a right eigen resolution of the identity (\ref{threee}), where 
\begin{equation*}
\Pi \left( \mathrm{d}\varkappa \right) =\int \delta \left( \mathrm{d}%
\varkappa \,,\varphi \left( \alpha \right) \right) \left\vert \alpha
\right\rangle \left\langle \alpha \right\vert \prod_{i=1}^{r}\frac{1}{\pi }%
\mathrm{d}\func{Re}\,\alpha _{i}\mathrm{d}\func{Im}\alpha _{i}
\end{equation*}%
($\delta \left( \mathrm{d}\varkappa ,\lambda \right) $ is the Dirac delta
measure of unite mass at the point $\lambda $). Therefore, the optimal
estimation of the parameters $\vartheta _{k}=\partial \ln \chi /\partial 
\bar{\beta}^{k}$ of the density operators (\ref{twoa}) for $\hat{x}=\varphi
\left( \hat{a}\right) $ is right efficient and reduces to a coherent
measurement and extension of the estimate $\vartheta =\varphi \left( \alpha
\right) $ with respect to the result $\alpha $. For the special case when
the function $\varphi \left( \alpha \right) $ is linear and the state $%
\varrho _{0}$ is Gaussian, this fact was established in \cite{5}.

Note that besides right and left lower bounds one can also consider other,
combined bounds by means of the factorization \cite{10} $\vartheta
=\vartheta _{+}+\vartheta _{-}$, defining right derivatives with respect to $%
\vartheta _{+}$ and left derivatives with respect to $\vartheta _{-}$. An
interesting question is this: Is the class of efficient statistics exhausted
by statistics for which at least one such bound can be attained?

\textbf{4. }In conclusion, let us consider the question of the (right)
efficiency of the estimation of the parameters $\beta ^{k}$ themselves of
the canonical families (\ref{twoa}). The inequality (\ref{i}) corresponding
to this case $\vartheta ^{k}=\beta ^{k}$ has the form $\mathsf{R}\geq 
\mathsf{H}^{-1}$ where $\mathsf{H}$ is the matrix of the derivatives (\ref%
{twoe}). Without loss of generality, we shall assume that $\mathrm{Tr}\hat{x}%
_{k}\varrho _{0}=0$.

\begin{theorem}
\label{Tthree}The inequality $\mathsf{R}\geq \mathsf{H}^{-1}$ becomes an
equality if and only if the operators $\hat{x}_{k}$ in (\ref{twoa}) have a
right joint spectral measure (\ref{threee}), the generating function of the
moments (\ref{twob}) of these operators in the state $\varrho _{0}$ is
Gaussian: $\chi \left( \beta ,\bar{\beta}\right) =\exp \left\{ \bar{\beta}%
^{i}H_{ik}\beta ^{k}\right\} $, where $H_{ik}$ does not depend on $\beta $
and $\bar{\beta}$, and the unbiased estimates $\lambda ^{k}=\lambda
^{k}\left( \varkappa \right) $ are taken to be linear functions $\lambda
^{k}=H^{ki}\varkappa _{i}$ of the results $\varkappa _{k}$ of simultaneous
quasimeasurement of the observables $\hat{x}_{k}$.
\end{theorem}

The proof of the sufficiency of these conditions for the existence of the
right efficient estimation is obvious: From the fact that the matrix $%
\mathsf{H}$ coincides with the covariance matrix $\mathsf{S}$ of the
operators $\hat{x}_{k}$ it follows that the covariance matrix $\mathsf{R}=%
\mathsf{H}^{-1}\mathsf{HH}^{-1}$ is equal to $\mathsf{H}^{-1}$.

The necessity follows from the necessary conditions of right efficiency of
Theorem \ref{Ttwo}, according to which the family $\varrho \left( \beta ,%
\bar{\beta}\right) $ must also have the form 
\begin{subequations}
\begin{equation}
\varrho \left( \beta ,\bar{\beta}\right) =\chi ^{-1}\mathrm{e}^{\beta _{k}%
\hat{x}^{k\ast }}\varrho _{0}\mathrm{e}^{\bar{\beta}_{k}\hat{x}^{k}},
\label{threej}
\end{equation}%
where $\chi \left( \beta ,\bar{\beta}\right) =\mathrm{Tr}\varrho _{0}\mathrm{%
e}^{\beta _{k}\hat{x}^{k}}\mathrm{e}^{\bar{\beta}_{k}\hat{x}^{k\ast }}$, $%
\frac{\partial }{\partial \bar{\beta}_{k}}\ln \chi =\beta ^{k}$, and the
operators $\hat{x}^{k}$ have the joint right resolution of the identity 
\end{subequations}
\begin{equation*}
\hat{1}=\int \Pi \left( \mathrm{d}\varkappa \right) ,\quad \hat{x}^{k}\Pi
\left( \varkappa \right) =\varkappa ^{k}\Pi \left( \mathrm{d}\varkappa
\right) ,\quad \varkappa =\left( \varkappa ^{k}\right) \in \mathbb{C}^{n}.
\end{equation*}%
Comparing (\ref{threea}) and (\ref{twoa}), we obtain $\bar{\beta}_{k}\hat{x}%
^{k}=\bar{\beta}^{k}\hat{x}_{k}$, whence 
\begin{equation*}
\beta _{k}=H_{ki}\beta ^{i},\quad \chi \left( \beta ,\bar{\beta}\right) =%
\bar{\beta}^{i}H_{ik}\beta ^{k},\quad \hat{x}^{k}=H^{ki}\hat{x}_{i}.
\end{equation*}%
Theorem 3 has been proved.

\section{Appendix}

\textbf{1.} Let us proof the inequality (\ref{i}). First consider the
one-dimensional case. Let $\hat{q}$ be an operator in $\mathcal{H}$ for
which 
\begin{equation}
\mathrm{Tr}\hat{q}\varrho \left( \alpha ,\bar{\alpha}\right) =\vartheta
\left( \alpha ,\bar{\alpha}\right) .  \label{Aa}
\end{equation}%
Differentiating (\ref{Aa}) with respect to $\alpha $ and using the
definition (\ref{g}) and the normalization condition $\mathrm{Tr}\varrho
\left( \alpha ,\bar{\alpha}\right) =1$, due to which $\mathrm{Tr}\varrho 
\hat{h}^{\ast }=0$, we obtain 
\begin{equation*}
\frac{d\vartheta }{d\alpha }=\mathrm{Tr}\varrho \left( \hat{q}-\vartheta
\right) \hat{h}^{\ast }.
\end{equation*}%
Since the covariance $\mathrm{Tr}\varrho \left( \hat{q}-\vartheta \right) 
\hat{h}^{\ast }$ satisfies the Schwarz inequality 
\begin{equation}
\left\vert \mathrm{T}\left[ \mathrm{r}\varrho \left( \hat{q}-\vartheta
\right) \hat{h}^{\ast }\right] \right\vert ^{2}\leq \mathrm{Tr}\left[
\varrho \left( \hat{q}-\vartheta \right) \left( \hat{q}-\vartheta \right)
^{\ast }\right] \mathrm{Tr}\left[ \varrho \hat{h}\hat{h}^{\ast }\right] ,
\label{Ab}
\end{equation}%
which is the condition of non-negativity of the determinant of the $2\times
2 $ matrix of covariances $\mathrm{Tr}\varrho \hat{h}_{i}\hat{h}_{k}^{\ast }$%
, $i=0,1$, where $\hat{h}_{0}=\left( \hat{q}-\vartheta \right) ,\;\hat{h}%
_{1}=\hat{h}$, we can write 
\begin{equation}
\mathrm{Tr}\varrho \left( \hat{q}-\vartheta \right) \left( \hat{q}^{\ast }-%
\bar{\vartheta}\right) \geq \left. \left\vert \frac{d\vartheta }{d\alpha }%
\right\vert ^{2}\right/ \mathrm{Tr}\varrho \hat{h}\hat{h}^{\ast },
\label{Ac}
\end{equation}%
This inequality obviously establishes a lower bound for the variance of the
estimation of the parameter $\vartheta =\vartheta \left( \alpha ,\bar{\alpha}%
\right) $ in the class of ordinary measurements described by normal
operators $\hat{q}$. However since the normality condition $\hat{q}\hat{q}%
^{\ast }=\hat{q}^{\ast }\hat{q}$ was not used in the derivation of (\ref{Ac}%
), this bound gives a lower bound for the variance of any unbiased
estimation of $\vartheta $. Indeed, if $\Pi \left( \mathrm{d}\lambda \right)
,\lambda \in \mathbb{C}$ is a QPM describing the unbiased estimation as a
generalized measurement in $\mathcal{H}$, then the operator $\hat{q}=\int
\lambda \Pi \left( \mathrm{d}\lambda \right) $ satisfy the condition (\ref%
{Aa}). From the Hermitian positivity 
\begin{equation}
\left( \lambda -\hat{q}\right) \Pi \left( \mathrm{d}\lambda \right) \left(
\lambda -\hat{q}\right) ^{\ast }\geq 0\quad \left( \Pi \geq 0\right)
\label{Ad}
\end{equation}%
it follows that $\int \left\vert \lambda \right\vert ^{2}\Pi \left( \mathrm{d%
}\lambda \right) \geq \hat{q}\hat{q}^{\ast }$, and 
\begin{equation}
\int \left\vert \lambda -\hat{q}\right\vert ^{2}\Pi \left( \mathrm{d}\lambda
\right) \geq \left( \lambda -\hat{q}\right) \left( \lambda -\hat{q}\right)
^{\ast }.  \label{Ae}
\end{equation}%
Taking the mathematical expectation of both sides of (\ref{Ad}) and bearing
in mind that the variance $R=\mathsf{M}_{\vartheta }\left[ \left\vert
\lambda -\vartheta \right\vert ^{2}\right] $ of the estimation $\vartheta $
is 
\begin{equation*}
R=\mathrm{Tr}\varrho \int \left\vert \lambda -\vartheta \right\vert ^{2}\Pi
\left( \mathrm{d}\lambda \right) ,
\end{equation*}%
we obtain in conjunction with (\ref{Ac}) 
\begin{equation}
R\geq \mathrm{Tr}\varrho \left( \hat{q}-\vartheta \right) \left( \hat{q}%
-\vartheta \right) ^{\ast }\geq \left\vert d\right\vert ^{2}/g,  \label{Af}
\end{equation}%
where we have denoted $d=d\vartheta /d\alpha ,\;g=\mathrm{Tr}\varrho \hat{h}%
\hat{h}^{\ast }$. Thus, for the one-dimensional case the inequality (\ref{i}%
) has been proved.

\textbf{2.} Equality can be attained in (\ref{Ae}) if, first, the
expectations of the two sides of (\ref{Ae}) coincide and, second, the
Schwarz inequality becomes an equality. The first condition actually
establishes equality in (\ref{Ad}). More precisely:

\begin{lemma}
Suppose the ranges $\varrho \left( \alpha ,\bar{\alpha}\right) \mathcal{H}$
of density operators $\left\{ \varrho \left( \alpha ,\bar{\alpha}\right)
:\alpha \in \mathcal{O}\right\} $ generate the whole of $\mathcal{H}$. Then
the equality $\mathrm{Tr}\varrho R=0$ for any non-negative definite operator 
$R$ in $\mathcal{H}$ and all $\alpha \in \mathcal{O}$ implies that $R=0$.
\end{lemma}

It is sufficient to show that in $\mathcal{H}$ there is no vector $%
\left\vert \chi \right\rangle $ of the form $\left\vert \chi \right\rangle
=\varrho ^{1/2}\left\vert \psi \right\rangle $ for which $\left\langle \chi
\right\vert R\left\vert \chi \right\rangle \neq 0$. But this follows from
the inequality 
\begin{equation*}
\mathrm{Tr}\varrho ^{1/2}R\varrho ^{1/2}\geq \left\langle \psi \right\vert
\varrho ^{1/2}R\varrho ^{1/2}\left\vert \psi \right\rangle .
\end{equation*}%
which holds for any non-negative $R$ when $\left\langle \psi \mid \psi
\right\rangle =1$.

Applying this result to the operator $R$ equal to the difference of the
right- and left-hand sides of (\ref{Ae}), we find, under the conditions of
the lemma, that equality holds in (\ref{Ae}) only if 
\begin{equation*}
\left( \lambda -\hat{q}\right) \Pi \left( \mathrm{d}\lambda \right) \left(
\lambda -\hat{q}\right) ^{\ast }=0\text{, or }\hat{q}\Pi \left( \mathrm{d}%
\lambda \right) =\lambda \Pi \left( \mathrm{d}\lambda \right) .
\end{equation*}

This proves that for the existence of right efficient unbiased estimation in
some domain $\mathcal{O}\ni \alpha $ it is necessary to have an operator $%
\hat{q}$ with a right-eigen QPM in the subspace generated by the subspaces $%
\varrho \left( \alpha ,\bar{\alpha}\right) \mathcal{H}$, with $\mathrm{Tr}%
\hat{q}\varrho \left( \alpha ,\bar{\alpha}\right) =\lambda $. In the case of
real spectrum $\lambda \in \mathbb{R}$ such an operator $\hat{q}$ is
obviously selfadjoint.

The second condition of equality in (\ref{Af}) is equivalent to the
condition of linear dependence $\varrho \left( \hat{q}-\lambda \right)
=t\varrho \hat{h}$, where $t=d/g$ is a constant. Setting 
\begin{equation*}
t\hat{s}=\hat{q}-\vartheta \left( 0\right)
\end{equation*}%
we obtain the equations 
\begin{equation*}
\partial \varrho /\partial \bar{\alpha}=\varrho \left( \hat{s}-\mu \right)
,\quad \partial \varrho /\partial \alpha =\left( \hat{s}-\mu \right) ^{\ast
}\varrho
\end{equation*}%
where $t\mu =\vartheta \left( \alpha \right) -\vartheta \left( 0\right) $.
Its solution of these equations with the boundary condition $\varrho \left(
0,0\right) =\varrho _{0}$ has the canonical form (\ref{twoa}). The the
operator $\hat{q}=t\hat{s}+\vartheta \left( 0\right) $ should have
right-eigen QPM, so the operator $\hat{s}$ should. This proves for the
one-dimensional case, the necessity of the canonicity of the density
operators $\varrho \left( \alpha ,\bar{\alpha}\right) $ for the existence of
the right efficient estimation formulated in Theorem 2. In the Hermitian
case $\hat{x}^{\ast }=\hat{x}$, this also proves the necessity of Theorem 1.

\textbf{3.} A multidimensional generalization is obtained from the
one-dimensional case by taking 
\begin{equation*}
\hat{q}-\lambda =\left( \hat{q}_{i}-\lambda _{i}\right) \bar{\eta}^{i},\quad 
\hat{h}=\hat{h}_{k}\bar{\xi}^{k},
\end{equation*}%
where $\eta ^{i},\;i=1,\ldots ,m$, $\alpha ^{k},\;k=1,\ldots ,n$, are
arbitrary complex numbers. Remembering that then 
\begin{equation*}
\mathrm{Tr}\varrho \left( \hat{q}-\lambda \right) \hat{h}^{\ast }=\bar{\eta}%
^{i}\frac{\partial \vartheta _{i}}{\partial \alpha ^{k}}\xi ^{k},
\end{equation*}%
we obtain from (\ref{Ab}) for $\xi ^{k}=\left( \mathsf{H}^{-1}\mathsf{D}%
^{\dagger }\right) _{i}^{k}\eta ^{i}$ the second of the inequalities 
\begin{equation*}
R_{ik}\bar{\eta}^{i}\eta ^{k}\geq \mathrm{Tr}\varrho \left( \hat{q}%
_{i}-\lambda _{i}\right) \left( \hat{q}_{k}-\lambda _{k}\right) ^{\ast }\bar{%
\eta}^{i}\eta ^{k}\geq \left( \mathsf{DH}^{-1}\mathsf{D}^{\dagger }\right)
_{ik}\bar{\eta}^{i}\eta ^{k},
\end{equation*}%
which holds for arbitrary $\hat{q}_{i}$ for which $\mathrm{Tr}\varrho \hat{q}%
_{i}=\vartheta _{i}$. Setting 
\begin{equation*}
\hat{q}_{i}=\int \lambda _{i}\Pi \left( \mathrm{d}\lambda \right) \text{,
where }\int \Pi \left( \mathrm{d}\lambda \right) =\hat{1},\;\lambda \in 
\mathbb{C}^{m},
\end{equation*}%
is the resolution of the identity describing the estimator $\lambda
_{i}=\varkappa _{i}$, and applying the inequality (\ref{Ae}) for $\hat{q}=%
\hat{q}_{i}\bar{\eta}^{i},\;\lambda =\lambda _{i}\bar{\eta}^{i}$, we obtain
for the matrix $\mathsf{R}$ of the covariances of $\vartheta _{i}$
satisfying the first of the inequalities (\ref{Af}), whence (\ref{i})
follows because $\eta ^{i}$ is arbitrary.

The inequality (\ref{h}) becomes an equality for $\alpha \in \mathcal{O}$
only if 
\begin{equation*}
\hat{q}_{i}\Pi \left( \mathrm{d}\lambda \right) =\lambda _{i}\Pi \left( 
\mathrm{d}\lambda \right) ,\text{ and }\varrho \left( \hat{q}_{i}-\lambda
_{i}\right) =t_{k}^{i}\partial \varrho /\partial \bar{\alpha}_{k},\text{
where }t_{k}^{i}=\left( \mathsf{DH}^{-1}\right) _{k}^{i},
\end{equation*}%
whence with allowance for $\mathsf{T}=\left[ T_{k}^{i}\right] $ to be
constants nondegenerated matrix we obtain (\ref{twoa}).

\end{document}